\documentclass{aa}
\usepackage{graphicx}
\usepackage{txfonts}

\def\HI{{\ion{H}{I}}}
\def\OIII{{[\ion{O}{III}]}}
\def\HeII{{\ion{He}{II}}}

\begin{document}

\title{The jet-ISM interaction in the outer filament of Centaurus A}

\author{F. Santoro \inst{1,2}\fnmsep\thanks{email: santoro@astro.rug.nl},
             J. B. R. Oonk\inst{1,3},
             R. Morganti\inst{1,2},
             and T. Oosterloo\inst{1,2} }

\institute{ASTRON, Netherlands Institute for Radio Astronomy, PO 2, 7990 AA, Dwingeloo, Netherlands.\and Kapteyn Astronomical Institute, University of Groningen, PO 800, 9700 AV Groningen, Netherlands.\and Leiden Observatory, Leiden University, PO Box 9513, 2300 RA Leiden, the Netherlands.}

\date{Received 03/10/2014; accepted 17/11/2014}
 
\abstract 
   {The interaction between the radio plasma ejected by the active nucleus of a galaxy and the surrounding medium is a key process that can have a strong impact on the interstellar medium of the galaxy and hence on galaxy evolution.
The closest laboratory where we can observe and investigate this phenomenon is the radio galaxy Centaurus A. About 15 kpc northeast of this galaxy, a particularly complex region is found: the so-called outer filament where it has been proposed that jet-cloud interactions occur. }
   {We investigate the signatures of jet-ISM interaction by a detailed study of the kinematics of the ionized gas, expanding on previous results obtained from the \HI.}
   {We observed two regions of the outer filament with VLT/Visible MultiObject Spectrograph (VIMOS) in Integral Field Unit (IFU) observing mode. Emission from H$\beta$ and \OIII$\lambda\lambda$4959,5007\AA\ is detected in both pointings.}
   {We found two distinct kinematical components of ionized gas that closely match the kinematics of the nearby \HI\ cloud. One component follows the regular kinematics of the rotating gas, while the second shows similar velocities to those of the nearby \HI\ component thought to be disturbed by an interaction with the radio jet.} 
   {We suggest that the ionized and atomic gas are part of the same dynamical gas structure stemming from the merger that shaped Centaurus A. It is regularly rotating around Centaurus A, as proposed by other authors.  The gas (ionized and \HI) with anomalous velocities traces the interaction of the large-scale radio jet with the interstellar medium, suggesting that the jet is still active although poorly collimated. However, we can exclude that a strong shock is driving the ionization of the gas.  
   It is likely that a combination of jet entrainment and  photoionization by the UV continuum from the central engine is needed to explain both the ionization and the kinematics of the gas in the outer filament. }


\maketitle
%

\section{Introduction}

The ejection of radio plasma jets is one of the main manifestations of nuclear activity in early-type galaxies. The jets are known to influence the conditions of the surrounding interstellar medium (ISM). Signatures of this are seen at both large \citep{2009arXiv0906.2900N, 2011MNRAS.411.1641E, 2012NJPh...14e5023M} and small \citep{ 2010ApJ...716..131R,2010ApJ...711L..94R,2013Sci...341.1082M} scales in accordance with simulations \citep{2002A&A...395L..13M,2004ApJ...604...74F,2011ApJ...728...29W}. Recent studies of the interaction between a jet and its surrounding ISM have brought a number of surprises. One is that through such interactions {\sl cold} gas (atomic and molecular) can be accelerated to high velocities, producing massive gas outflows \citep{2010A&A...518L.155F,2012A&A...541L...7D,2013Sci...341.1082M,2014A&A...562A..21C,2014Natur.511..440T}. Compression of  gas by the jet can also have implications for jet-induced star formation \citep{1993ApJ...414..563V,1997ApJ...490..698D,1998ApJ...502..245G,2000ApJ...536..266M,2005A&A...429..469O}.

So far, only the most spectacular examples of these phenomena have been studied in detail, while, an interaction between a jet and the ISM is a relatively common phenomenon, also in low-power radio sources (\citealp[e.g., NGC~1266,][]{2011ApJ...735...88A}; \citealp[NGC~1433,][]{2013A&A...558A.124C}). Thus, understanding the occurrence, characteristics, and effects of such interactions is relevant to the broader question of the role of an active galactic nucleus (AGN) in galaxy evolution. 

Among the objects where signatures of jet-ISM interaction have been found is also the nearest AGN, Centaurus A (Cen~A), which is the target of the present study. This object combines all the ingredients that help trace such an interaction; in addition to the large and complex radio structure, possibly the result of multiple events, it  also  has a rich ISM.
Cen~A is the closest low-luminosity (FR~I) radio galaxy \citep[$d=3.8$ Mpc;][]{2010PASA...27..457H}. Its host galaxy (NGC~5128) is the dominant member of a poor group, and it appears to be in the late stages of a merger with a small late-type galaxy. A system of faint optical shells \citep{1983ApJ...272L...5M}, a stellar stream \citep{2002AJ....124.3144P}, and  \HI\ shell-like structures at large radii \citep{1994ApJ...423L.101S} have been found and are believed to be linked to this merger event.

\begin{figure*}
\centering
\includegraphics[width=13cm,keepaspectratio]{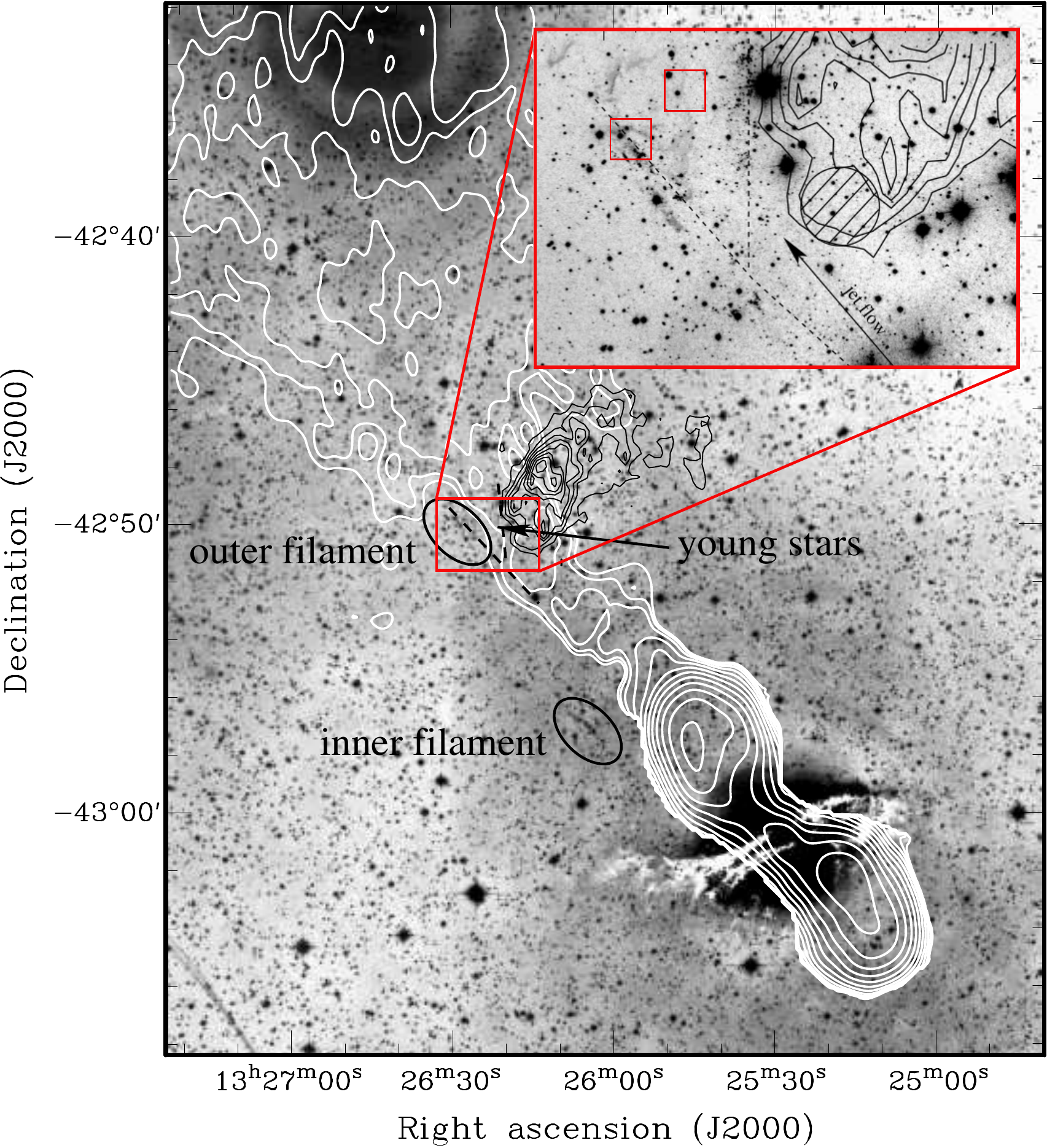}
      \caption{ Optical image of Cen~A (kindly provided by D. Malin) showing the diffuse emission and the location of the so-called inner and the outer filaments. The white contours denote the radio continuum emission of the inner lobes (bottom-right of the figure) and the large-scale jet connecting the northern inner Lobe to the base of the outer lobe (top-left of the figure). The black contours refer to the northeast outer \HI\ cloud. The location where young stars have been found is marked with a dashed line. \textit{Zoom in:} Optical image of the outer filament region (kindly provided by M. Rejkuba). The hatched area indicates  where anomalous velocities are detected within the \HI\ cloud. The region covered by our observations is drawn with two red squares.} 
\label{CenA}
\end{figure*}

The radio source shows a set of lobes possibly connected with different phases of restarted AGN activity \citep{1983PASAu...5..241H,1999MNRAS.307..750M}.
The so-called outer radio lobes are believed to be the result of a previous episode of nuclear activity, but they are possibly still supplied with fresh electrons via a poorly collimated jet-like structure, at least in the case of the northern lobe \citep[the large-scale jet,][]{1999MNRAS.307..750M,2009MNRAS.393.1041H}, while the inner radio lobes are the result of a recent outburst \citep{2009MNRAS.395.1999C}.  
The layout of the inner and outer lobes and the structure connecting the two is shown in Fig.~\ref{CenA}. 

Filaments of highly ionized gas are found between the inner radio lobe and the base of the outer lobe \citep{1975ApJ...198L..63B,1981ApJ...247..813G,2003AAS...203.9607N}. They appear quite well aligned along the direction of the inner jet. The best-studied ones are the two filamentary structures called the inner and outer filaments, which are at distances of 8.5 kpc and 15 kpc, respectively, from the nucleus.
Both filaments have been identified as regions where an interaction between the radio jet and gas clouds is occurring.

Particularly complex is the situation in the outer filament (see  Fig.~\ref{CenA}), and this has motivated the present study. In projection, this filament is located  close to the axis of the jet-like structure connecting inner and outer radio lobes, while very near the outer filament, a large \HI\ cloud is also found.  This cloud is part of a larger rotating \HI\ structure that appears to form a polar ring surrounding Cen~A \citep{1994ApJ...423L.101S}.  Evidence for anomalous velocities in the southern tip of the \HI\ cloud has been reported by \citet{2005A&A...429..469O}. Considering that this part of the \HI\ cloud is closest to the jet axis, this finding was interpreted  as this being the location where an interaction between the radio jet and the neutral gas takes place. The interaction can cause some gas to be dragged along the jet and displace it suggesting also a connection between the neutral gas and the ionized gas.  Importantly, regions of recent star formation \citep{1998ApJ...502..245G,2000ApJ...538..594F,2000ApJ...536..266M,2002ApJ...564..688R}, as well as young stars (as young as 15 Myr), have also been found in between the ionized and the neutral gas. The young stars are found to be distributed in a north-south direction along the eastern edge of the \HI\ cloud and may be the result of jet-induced star formation. 
\citet{2009ApJ...698.2036K} detected X-ray emission along the radio jet connecting the inner and outer lobes. The authors claim that cold gas is shock heated by a direct interaction with the jet, confirming that indeed the radio jet in this region could be an active channel through which energy is transferred from the center of the galaxy to the outer regions. 

It is clear from the above that the outer filament and its surroundings show a variety of complex features, all suggesting it is a region where the effects of a jet-cloud interaction are visible.
The aim of the observations presented here is to further study such an interaction and trace the characteristics and physical state of the gas involved in this process. We focus on the study of the ionized gas which has been shown earlier  to have a complex kinematics \citep{1991MNRAS.249...91M}.

\section{Data reduction and analysis}

We observed two fields in the outer filament using the Visible MultiObject Spectrograph \citep[VIMOS,][]{2003SPIE.4841.1670L} at the Very Large Telescope (VLT). In the following we refer to the northern field as Field 1 and the southern one as Field 2 (see  Fig.~\ref{TotalFluxMap}). 
The observations were done in Integral Field Unit (IFU) mode using the high-resolution blue grism covering the spectral range of about 4000-6200 \AA. The spectral resolution is $\sim$2.3 \AA\ corresponding to a velocity resolution ranging from 115 to 172 km s$^{-1}$.  Observations were carried out on March 10, 2006 with an exposure time of 3$\times$550 sec for each of the two fields. Each field covers about $27^{\prime\prime}\times27^{\prime\prime}$ and $\rm{1 pixel=0.67 arcsec \times 0.67 arcsec}$. The spatial resolution is limited by the seeing which is $\sim$ 1.4 arcsec. For flux calibration purposes, we used standard star calibration exposures taken during the same night.
The data reduction is carried out using $\rm{\small{P}3\small{D}}$, a data reduction tool optimized for IFU data \citep{2010A&A...515A..35S}. The reduction mainly involves five relevant steps: 1) the creation of a trace mask in order to locate the spectra on the CCD; 2) the extraction of a flat-field mask, to take into account the sensitivity variations of the detector; 3) the creation of a dispersion mask to perform the wavelength calibration; 4) spectra extraction; and 5) the creation of an overall sensitivity function to flux calibrate the scientific exposures. The average wavelength calibration accuracy is 0.1 \AA\ corresponding to $\sim$ 5 km s$^{-1}$. The flux calibration accuracy is $\sim$10$\%$.  

Ionized gas emission is clearly detected in the final datacube, and its spatial distribution appears consistent with \citet{1991MNRAS.249...91M}. H$\beta$ and \OIII$\lambda\lambda$4959,5007\AA\ emission is detected in both fields.
The bright \OIII$\lambda$5007\AA\ line is used to display the distribution and the kinematics of the ionized gas. The fainter H$\beta$ emission is found to be consistent with the \OIII$\lambda$5007\AA\ both in spatial and velocity distribution. 

In Field 2, at many positions two distinct velocity components are observed.
To parametrize this,  two simple models are employed for the line fitting: a single gaussian and a double gaussian model. For all gaussians, we fix the full-width-at-half-maximum (FWHM) to the instrumental spectral resolution.
Both models are fit to each spectrum and a $\chi^{2}$ statistic test is used to determine the best model. 
For the regions of blending, it has been verified that our double gaussian model with fixed FWHM is always better than a single gaussian model with the line width as a free parameter. We conclude that the intrinsic line width of the components is less then the instrumental spectral resolution.

The \OIII$\lambda$5007/H$\beta$ line ratios are extracted across both fields. Due to the fainter H$\beta$ emission and the low S/N we are not able to show line ratio maps on a pixel to pixel basis.
The velocity and flux maps are extracted using spectra for which the \OIII$\lambda$5007\AA\ emission is detected with a S/N$\geq$5. 

\section{Results}

\begin{figure}[t]
\centering
\includegraphics[width=\hsize, keepaspectratio]{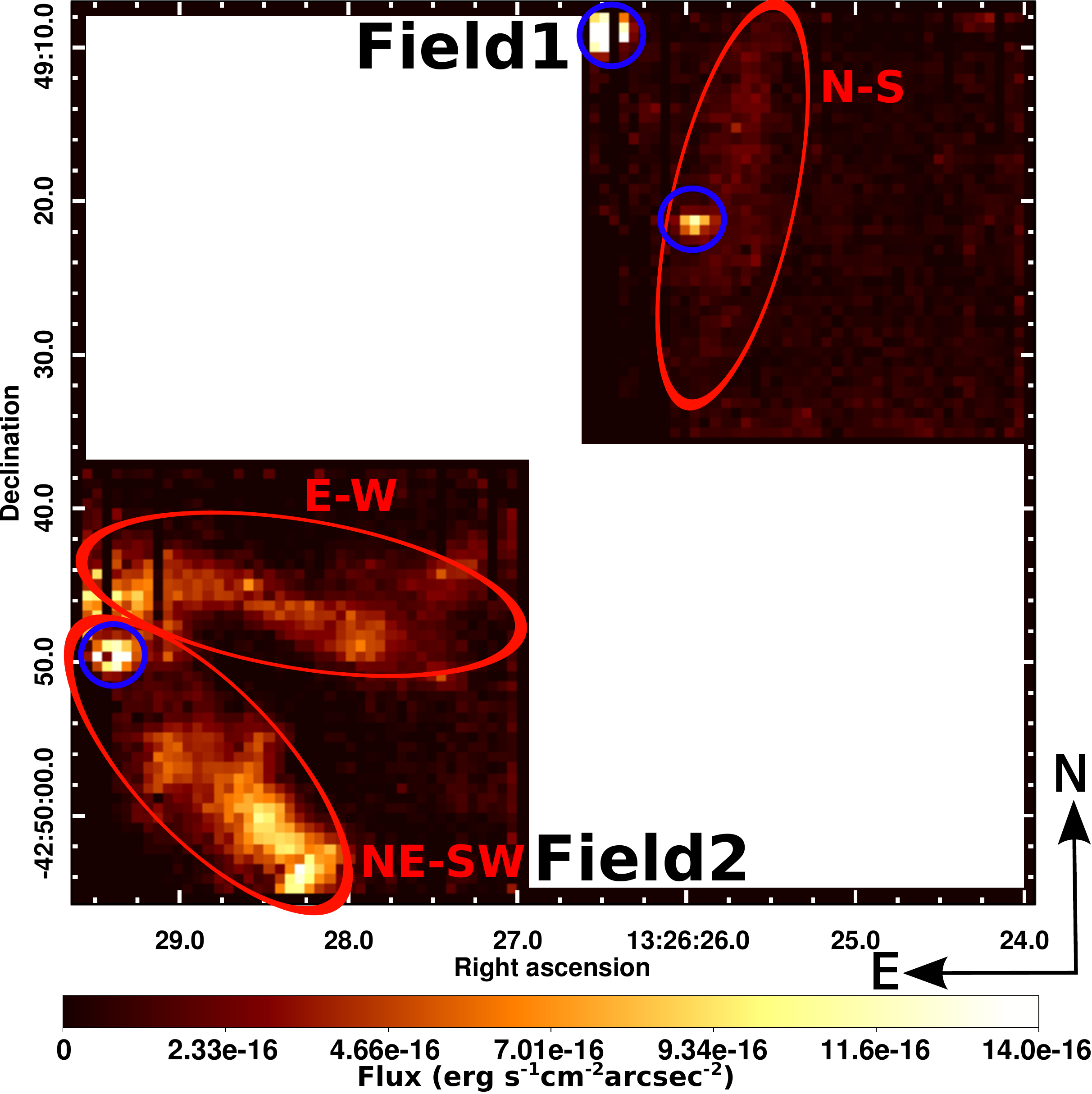}
\caption{Total \OIII$\lambda$5007\AA\ line flux map. Stronger ionized gas emission is visible in Field 2 while fainter emission is present in Field 1. Looking at the spatial distribution of the ionized gas, we locate three main filamentary structures that are outlined using red ellipses. Blue circles mark emission related to stars or associations of stars \citep[see][]{2002ApJ...564..688R}.}
\label{TotalFluxMap}
\end{figure}

The distribution of the ionized gas detected in our observations is consistent with what found in the narrow band images presented in Morganti et al. (1991), although we sample only a small part of the filament. The Fig.~\ref{TotalFluxMap} shows the distribution of the total \OIII$\lambda$5007\AA\ emission across the observed fields. 
The ionized gas is distributed in clumpy structures across both our fields. 
Looking at the overall gas distribution in Fig.~\ref{TotalFluxMap}, three main structures can be identified. Two of them are in Field 2, extending northeast-southwest and east-west respectively, while another one is in Field 1, extending north-south. These structures are elongated, with an extent of at least 20$^{\prime\prime}$ in length corresponding to $ \sim $0.37 kpc. 
The velocity range covered by the ionized gas (200-450 km s$^{-1}$) is found to be consistent with what found from long slit data by \citet{1991MNRAS.249...91M}.
However, the main finding of our observations is that we detect and isolate two kinematical components of the ionized gas. The components are, in part, spatially superimposed. 

\begin{figure}[t]
\centering
\includegraphics[width=\hsize, keepaspectratio]{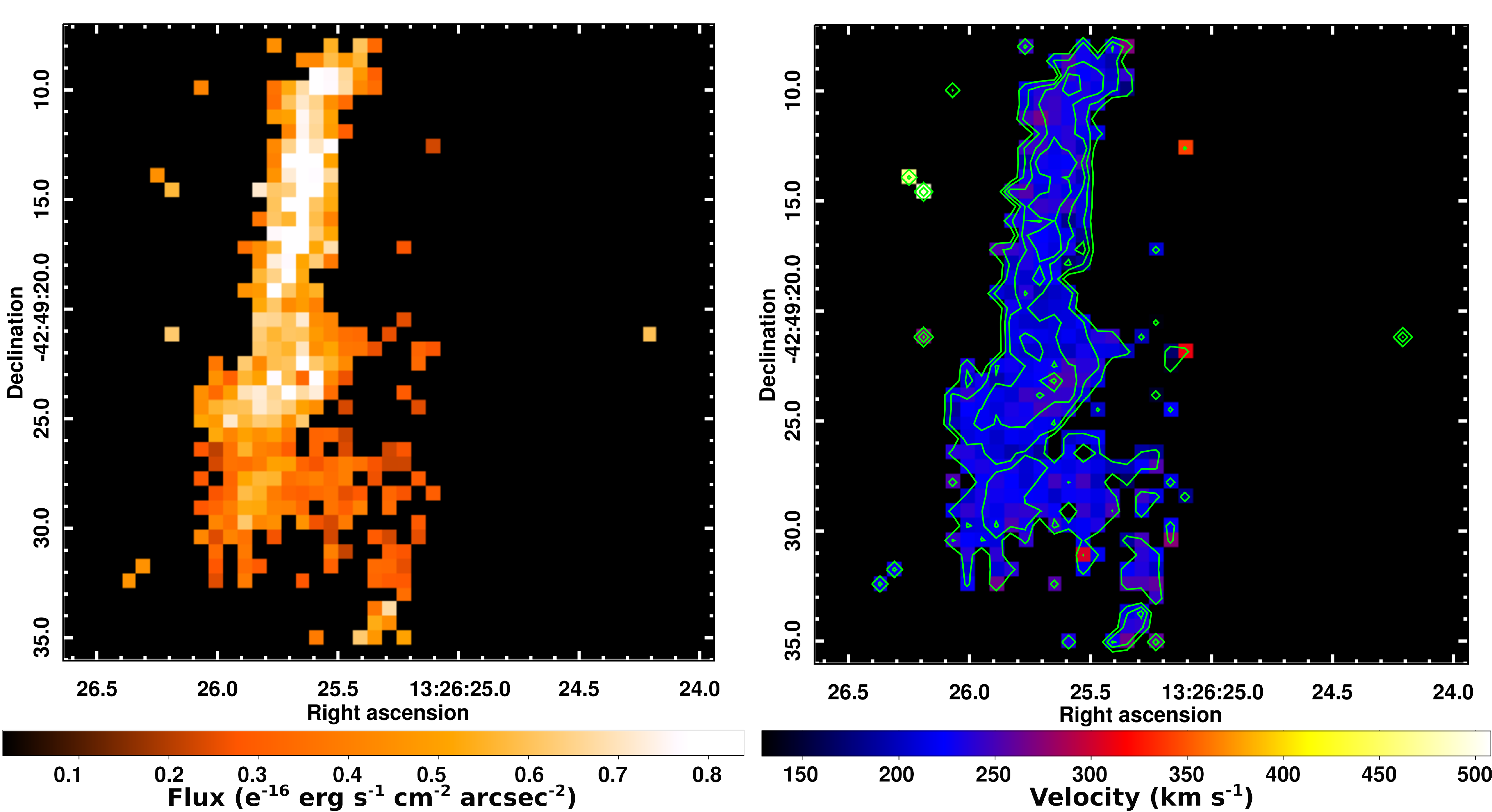}
\caption{Total \OIII$\lambda$5007\AA\ line flux (left panel) and velocity (right panel) map for the first ionized gas component in the Field 1. Intensity contours are overplotted in green on the velocity map. Contour levels are 0.17, 0.51, 0.71, 0.88 $\times 10 ^{-16}{\rm erg~s ^{-1}cm ^{-2}arcsec ^{-2}}$.}
\label{comp1filed1}
\end{figure}

\begin{figure}[t]
\centering
\includegraphics[width=\hsize, keepaspectratio]{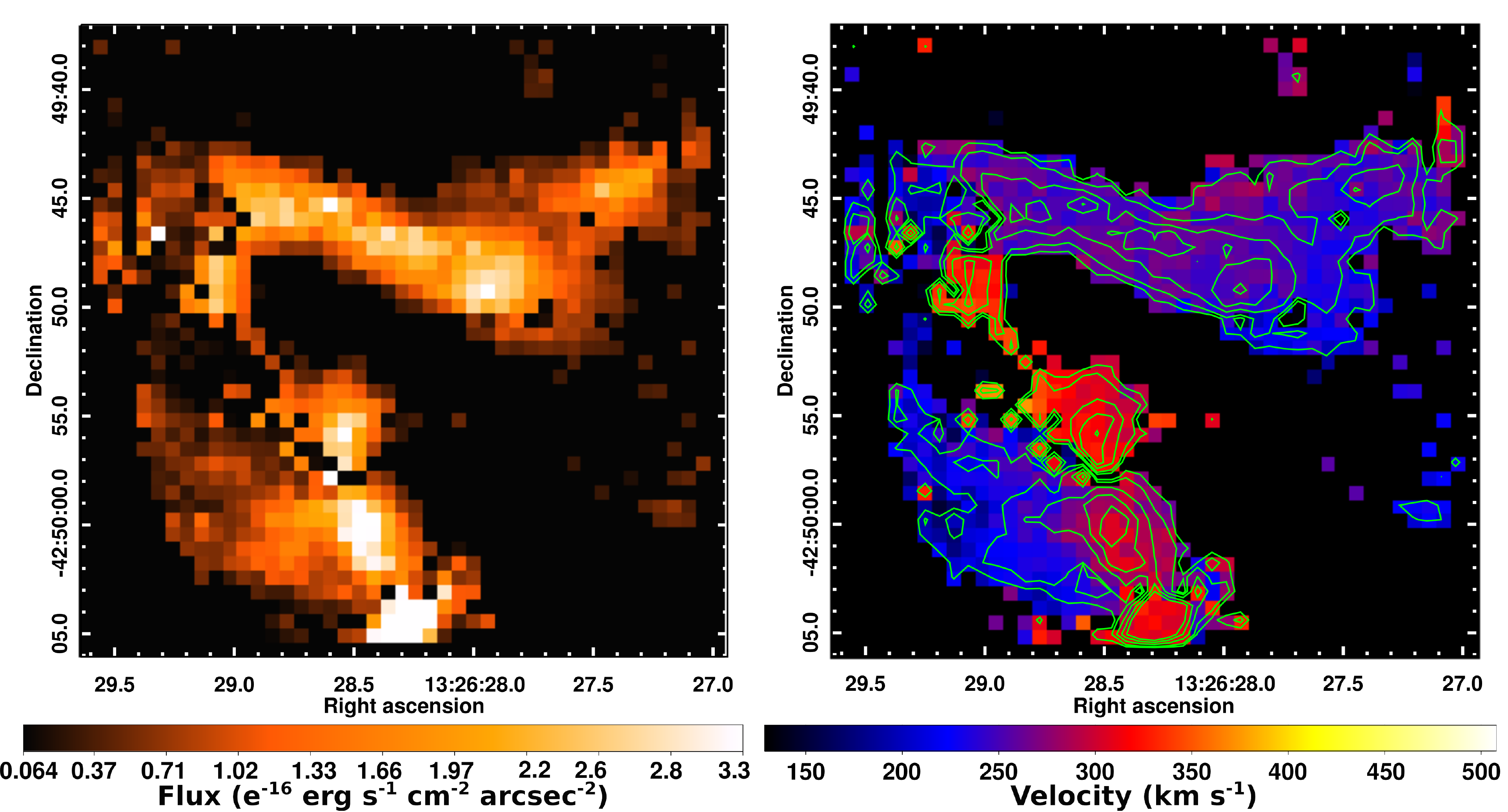}
\caption{Total \OIII$\lambda$5007\AA\ line flux (left panel) and velocity (right panel) map for the first ionized gas component in the Field 2. Intensity contours are overplotted in green on the velocity map. Contour levels are 0.4, 0.8, 1.6, 2.4, 3.1 $\times 10 ^{-16}{\rm erg~s ^{-1}cm ^{-2}arcsec ^{-2}}$.}
\label{comp1field2}
\end{figure}

\begin{figure}
\centering
\includegraphics[width=\hsize, keepaspectratio]{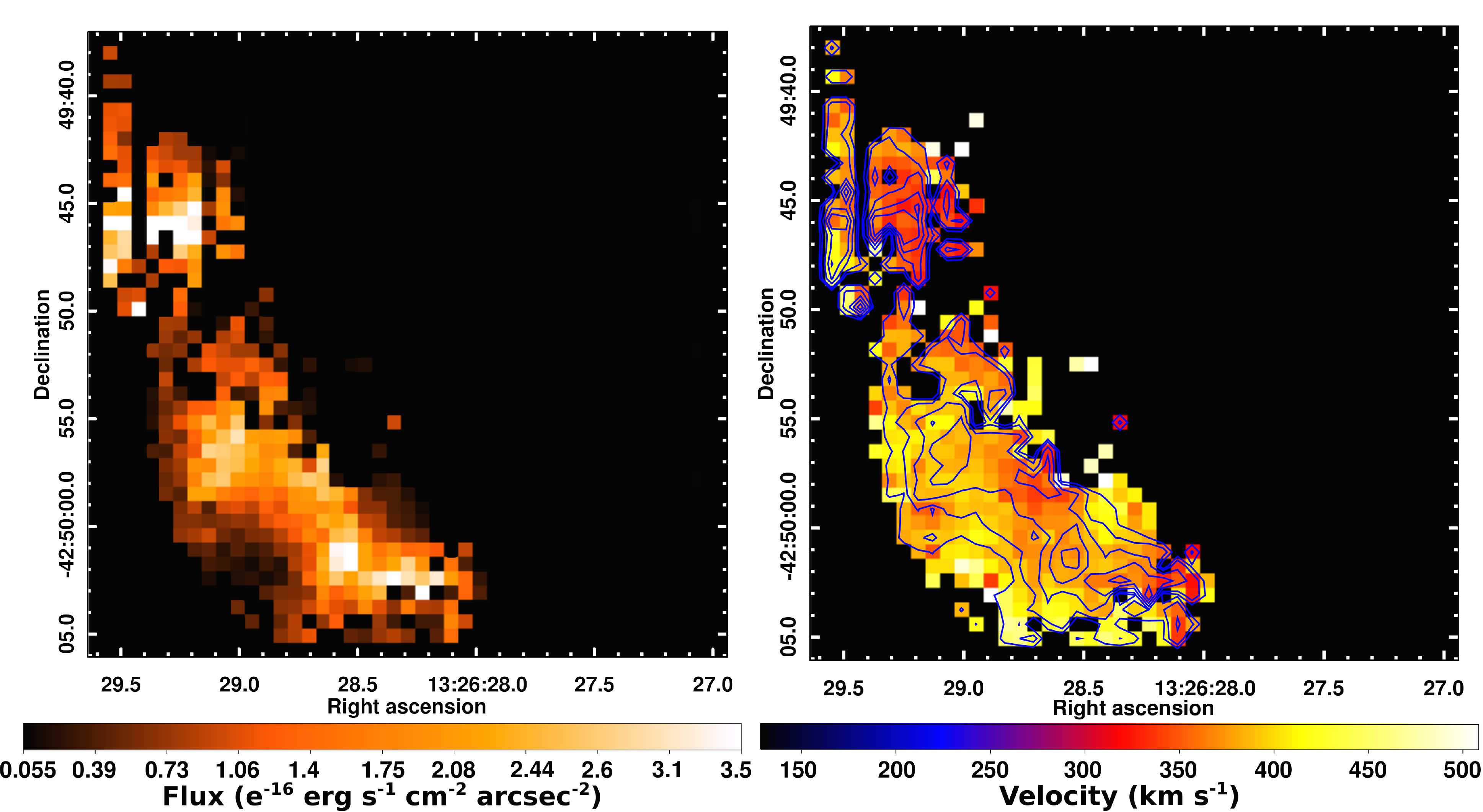}
\caption{Total \OIII$\lambda$5007\AA\ line flux (left panel) and velocity (right panel) map for the second ionized gas component that only extends across Field 2. Intensity contours are overplotted in blue on the velocity map. Contour levels are 0.4, 0.8, 1.6, 2.4, 3.1 $\times 10 ^{-16}{\rm erg~s ^{-1}cm ^{-2}arcsec ^{-2}}$.}
\label{comp2field2}
\end{figure}

The low-velocity component spreads across both  observed fields and has a mean heliocentric velocity of v$_{\rm hel}\sim$250 km s$^{-1}$. The Figures~\ref{comp1filed1} and ~\ref{comp1field2} show the \OIII$\lambda$5007\AA\ line flux and the velocity of the low-velocity component across both fields. The high-velocity component has a mean heliocentric velocity of v$_{\rm hel}\sim$400 km s$^{-1}$ and is only detected in Field 2. The Fig.~\ref{comp2field2} shows the corresponding \OIII$\lambda$5007\AA\ line flux and velocity map. 
   
\begin{figure}
\centering
\includegraphics[width=\hsize]{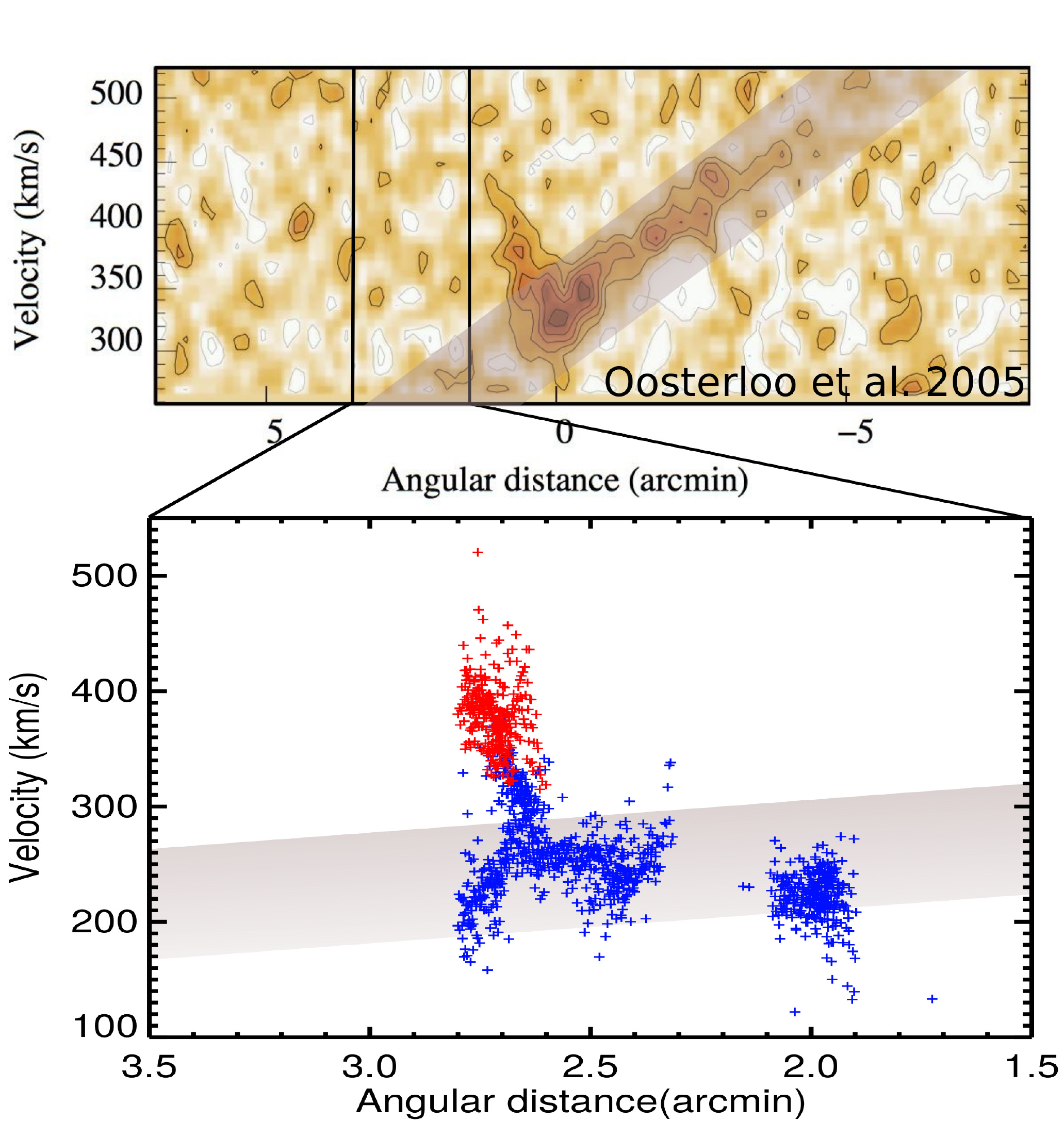}
\caption{Position-velocity plot of the nearby \HI\ cloud (upper panel) and of the ionized gas (lower panel). In the ionized gas position-velocity plot, the points related to the low and the high velocity components are drawn with blue and red crosses respectively. The shadowed stripe marks the large-scale velocity gradient of the \HI\ cloud. In order to calculate the angular distance, the zero point along the horizontal axis is assumed to be located at RA=13$^{h}$26$^{m}$15$^{s}$ and DEC=-42$^{\circ}$49$^{\prime}$00$^{\prime\prime}$ in accordance with the position-velocity plot shown in the upper panel.  }
\label{pvplot}
\end{figure}

An interesting question is how the velocity structure we observe for the ionized gas relates to that of the nearby \HI\ cloud. The overall kinematics of the \HI\ cloud is characterized by a smooth velocity gradient of about 30 km s$^{-1}$kpc$^{-1}$, with velocity decreasing when going east. This gradient corresponds to the rotation of the cloud around the galaxy \citep{1994ApJ...423L.101S}. In addition,  \citet{2005A&A...429..469O} found  the gas in the southern tip of the cloud to show a sudden velocity reversal. It should be noted that the \HI\ cloud is located at about 2.3 kpc distance from the ionized gas. However, the  kinematical features detected in the ionized gas appear to show a striking similarity with these  \HI\ velocities which is illustrated in Fig.~\ref{pvplot}. This plot has been obtained by extrapolating the velocity pattern of the regular rotating \HI\ from \citet{2005A&A...429..469O} (shown in the upper panel) and superposing it on our ionized gas velocity measurements (lower panel).

The first thing to notice is that the velocities of the low-velocity component, which is seen in both fields, falls on top of the extrapolation of the overall velocity gradient  of the regularly rotating \HI\ (shown in the upper panel of Fig.~\ref{pvplot}). It, therefore, appears that this component is an extension of the \HI\ cloud. The second striking feature is, however, that the velocities of the high-velocity component are very similar to the velocities of the 'anomalous' gas at the southern tip of the \HI\ cloud. It appears, therefore, that this anomalous  \HI\ has an ionized counterpart.

We have also used the results of the line fits of the ionized gas to investigate whether differences in ionization are present across the two fields. 
Average values of the extracted line ratios are reported in Table~\ref{table} for each of the three ionized gas structures. Within the area covered by our VIMOS observations there is no evidence of any strong spatial variations of the \OIII$\lambda$5007/H$\beta$ ratio. Due to the weakness of the H$\beta$ line the signal to noise is not sufficient to decompose this line in the northeast-southwest structure. Therefore in this filament the \OIII$\lambda$5007/H$\beta$ ratio, given in Table~\ref{table}, corresponds to total integrated emission for both velocity components. Our \OIII$\lambda$5007/H$\beta$ ratios are consistent with those measured by \citet{1991MNRAS.249...91M}. We also searched for emission from \OIII$\lambda$4363\AA\ and \HeII$\lambda$4686\AA\ lines, but neither was detected. The 3$\sigma$ upper limits relative to \OIII$\lambda$5007\AA\ and H$\beta$ are given in Table~\ref{table} .

\begin{table*}[t]
\centering 
\begin{tabular}{c c c c c}       
\hline              
\noalign{\smallskip}
 & \textbf{\OIII$\lambda$5007/H$\beta$} & \textbf{\OIII$\lambda$4363/$\lambda\lambda$(4959+5007)} & \textbf{\OIII$\lambda$4363/$\lambda$5007} & \textbf{\HeII$\lambda$4686/H$\beta$} \\    
\hline\hline
\noalign{\smallskip} 
   \textbf{NE-SW} & 5.8 (+0.45 -0.39) & $\leq$ 0.03 & $\leq$ 0.05 & $\leq$ 0.4\\      
   \textbf{E-W} & 5.8 (+0.67 -0.55) &  $\leq$ 0.07  & $\leq$ 0.1 & $\leq$ 0.65\\
   \textbf{N-S} & 5 (+1.22 -0.86) & $\leq$ 0.08    & $\leq$ 0.1 & $\leq$ 0.51\\
\hline 
\noalign{\smallskip}
\end{tabular}
\caption{Average values of the \OIII$\lambda$5007/H$\beta$ ratio and the \OIII$\lambda$4363/$\lambda\lambda$(4959+5007), \OIII$\lambda$4363/$\lambda$5007, \HeII$\lambda$4686/H$\beta$ ratio 3$\sigma$ upper limit related to the ionized gas structures as outlined in Fig.~\ref{TotalFluxMap}. }                                     
\label{table}
\end{table*}

\section{Discussion and conclusions}

The main result from our observations is that we detect two kinematical components in the ionized gas in the outer filament and that the velocities of these two components match those of the nearby \HI\ cloud. Both the regularly rotating \HI\ structure, as well as that part of the \HI\ which appears disturbed by the jet, have a counterpart in the ionized gas.   
This strongly suggests that the ionized and  neutral components are part of the same gas structure.    \citet{1994ApJ...423L.101S} suggested that the \HI\ cloud is a tidal feature originating from the merger, and is part of a ring-like structure rotating about Cen~A.  We argue that the ionized gas found here is also part of this structure but corresponds to that part of the tidal feature that is strongly affected by the radio jet. 
The fact that the region where young stars have been detected \citep{2000ApJ...536..266M,2002ApJ...564..688R} is in between the ionized and neutral gas would suggest that the star formation in this region has been induced by the jet.

Based on these findings, one can arrive at the following scenario.
As part of the polar ring, the \HI\ cloud is rotating along an orbit with an estimated inclination of 60-70 degrees and a position angle of the major axis of about 15 degrees  \citep{1994ApJ...423L.101S}. The orientation of the rotational plane of this polar ring is such that the eastern half is tilted toward us than the main galaxy body, and the western half is behind the optical galaxy. Relative to the systemic velocity of Cen~A (v$_{\rm sys}\sim$540 km s$^{-1}$), the \HI\ cloud is moving toward us and, given the geometry of the polar ring, is located on that part of the ring that is tilted toward us.
The northeast inner radio jet is pointing toward us, with an inclination of 50-80 degrees \citep{1996ApJ...466L..63J,1998AJ....115..960T} and this can be taken as an indicative value for the inclination of the large-scale jet. It is, therefore, conceivable that, in the recent past, a part of the \HI\ cloud, due to its rotation about the galaxy,  has entered the zone of influence of the large-scale radio jet.
 
If this is indeed the case,  the geometry is such that this structure has rotated into the zone of influence of the jet form the back, moving towards us.
Compared to the regularly rotating \HI\ tidal feature, both the disturbed \HI\ and the  high-velocity ionized gas are redshifted (of $\sim$100 km s$^{-1}$). The effect of the jet interaction will be to change the motion of a fraction of the \HI\  as well as of the ionized gas. 
Part of this motion will be along the jet, but if the interaction  is mainly in the form of entrainment by a turbulent cocoon surrounding the jet, an effect of the jet-cloud interaction will be a lateral motion with respect to the jet axis. 
Given the geometry described above (i.e., the gas entering the zone of influence of the jet from the back), one would indeed expect that such lateral motions are away from the observer.

The uniformity of the \OIII$\lambda$5007/H$\beta$ line ratios suggest the existence of a spatially extended ionization mechanism. A strong contribution to the ionization by stars can be ruled out upon combining our results with those of \citet{1991MNRAS.249...91M}. Pure shock models also do not provide a satisfactory explanation for the measured line ratios, although shock+precursor models cannot be ruled out \citep[e.g.,][]{1995ApJ...455..468D}. 
The narrow velocity width of the emission lines we observe is in line with this and lead us to exclude that a strong shock drives gas ionization.

Models of an entraining jet have been applied to the filaments of Cen~A by \citet{1993ApJ...414..510S}. We suggest that an entraining jet is likely to explain the gas kinematics. The models of \citet{1993ApJ...414..510S} may apply when there are bulk turbulent  velocities $\geq$ 200 km s$ ^{-1} $. In our case the degree of turbulence of the gas, deduced from the line width of the emission lines, is observed to be smaller.
Considering also the fairly diffuse nature of the radio jet near the outer filament, a gentle jet-cloud interaction, in the form of entrainment by a cocoon, is likely to be expected. 

It should be noted that the ionized gas is  quite well aligned with the position angle of the inner radio jet. We can assume that this orientation also defines the position angle of an UV radiation cone coming from the nucleus. \citet{1991MNRAS.249...91M,1992MNRAS.256P...1M} have investigated the possible photoionization of the outer filament by the radiation field of a nuclear blazar-like source. They concluded that this mechanism may indeed be responsible for the ionisation of the gas. This is consistent with the uniformity of the line ratios.

$~$ 
 
In summary, we find evidence for a single dynamical gas structure at $\sim$15 kpc distance from the center of Cen~A as a result of the last merger. This tidal cloud includes ionized (the outer filament) and neutral (the outer northeast \HI\ cloud) gas rotating about the galaxy. The fact that the kinematics of the \HI\ and the ionized gas match very well suggests that it is likely that the radio jet is interacting with these clouds. This shows us how the jet of a typical FR~I galaxy can interact with the ISM  even at scales of tens of kpc, affecting the kinematic and physical conditions.    

MUSE data covering a significant amount of the ionized gas in the area of the outer filament have been obtained recently. This will allow an extension of the present study and a more complete investigation of the physical and kinematical state of the ionized gas.

\begin{acknowledgements}
The research leading to these results has received funding from the European Research Council under the European Union's Seventh Framework Programme (FP/2007-2013) / ERC Advanced Grant RADIOLIFE-320745. Based on observations made with ESO Telescopes at the La Silla Paranal Observatory under programme 076.B-0059A  
\end{acknowledgements}

\bibliographystyle{aa}
\bibliography{biblio.bib}

\end{document}